# Macroscopic coherence of a single exciton state in a polydiacetylene organic quantum wire

F. Dubin[1], R. Melet, T. Barisien, R. Grousson, L. Legrand, , M. Schott & V. Voliotis[2]

*Institut des Nanosciences de Paris, CNRS UMR 7588, Université Pierre et Marie Curie et Université Denis Diderot, Campus Boucicaut, 140 rue de Lourmel, F-75015 Paris, France*

**Up to now, macroscopic quantum coherence has been evidenced in many-body systems such as superconductors, quantum liquids[1] and cold atoms condensates[2]. In semiconductors, the elementary excitations are excitons, i.e. electron-hole pairs bound by Coulomb interaction. In an ideal crystal, these quasi-particles are delocalized plane waves extending over the entire volume of the semiconductor, but in practice, any disorder destroys the large spatial coherence. Excitons have been claimed to form a macroscopic coherent state in a quasi two dimensional dense exciton gas[3] (see however Snoke et al[4]) or a condensate phase (microcavity polaritons[5,6]). Here we show that a single exciton state in an individual ordered conjugated polymer chain[7,8], exhibits macroscopic quantum spatial coherence reaching tens of microns, limited by the chain length. The spatial coherence of the *k=0* exciton state is demonstrated by selecting two spatially separated emitting regions of the chain and observing their interference.**

1 present address: Institut für Experimentalphysik, Universität Innsbruck, Technikerstrasse 25, A-6020 Innsbruck, Austria

2 also at Université Evry Val d'Essonne, Boulevard F. Mitterrand, 91025 Evry Cedex, France

Polydiacetylene (PDA) chains, form a model system to investigate electronic properties of conjugated polymers[9]. Their optical excitation, an exciton, is usually considered as a neutral excited state of the chain[10]. However, a one dimensional excitonic energy band is necessary to account for polydiacetylenes optical properties[7, 8]. Here, we demonstrate that these macromolecules are actually more than model conjugated polymers: the chains are almost perfect semiconductor quantum wires allowing macroscopic delocalization of excitons. In the experiments reported here, a single organic quantum wire, a PDA chain, is studied at low temperature by means of microscopic imaging spectroscopy (Fig. 1 a). The wire excitonic fluorescence, when imaged, shows large spatial extension over tens of μm. When overlapping the emission coming from two spatially well separated regions of a single chain, interference fringes are observed, demonstrating a macroscopic spatial coherence length of the radiative exciton.

PDA chains diluted in their monomer crystalline matrix are very long, highly regular polymer chains, isolated from one another. They can take either of two conformations, conventionnally named « blue » and « red » which correspond to two electronic configurations with identical molecular structures[11]. This work exclusively deals with « red » chains. Disorder is very small, the inhomogeneous broadening of the absorption or emission of an ensemble of red chains being less than 2 meV [11, 12]. These chains emit a strong resonance fluorescence around 2.28 eV (at 10 K) and several vibronic replicas which are two or more orders of magnitude weaker. The emission is polarized parallel to the chain axis. Microphotoluminescence experiments are possible thanks to high dilution and to the intense and narrow resonance emission, then a single isolated chain can be studied [12]. We have shown, that a single red chain has the characteristic properties of a one dimensional semiconducting quantum wire. Indeed, the optical excitation is a highly bound exciton [13], described by a one dimensional energy band with the predicted $1/\sqrt{E}$ behaviour of the density of states [7], leading to an exciton effective lifetime scaling with temperature like $\sqrt{T}$ [8]. The single chain fluorescence comes from the radiatively coupled exciton states near *k=0*. It is lorentzian with an homogeneous broadening $\Gamma_h$ governed by the interaction between the exciton and the three dimensional longitudinal acoustic phonons of the monomer surrounding the chain [7, 12] ($\Gamma_h \approx$ 500 µeV at 10 K, see Fig. 1 c).

For resonant excitation, when pumping directly the radiating $k \approx 0$ exciton states, the fluorescence is much weaker than the strong scattered laser light, making its detection difficult to achieve. Therefore, we choose to photocreate excitons by pumping within the homogeneous linewidth of the absorption strongest vibronic replica at 2.47 eV (so-called D-absorption line corresponding to the double-bound stretch optical phonon). However, the $k \approx 0$ exciton states are almost instantaneously created within a few tens of femtoseconds [14]. Figure 1.b shows a spectrally resolved single chain fluorescence image. The low excitation power ($\approx$ 1 µW) ensures that the probability of photocreating an exciton in the chain during its radiative lifetime, is less than one. Hereafter, we therefore deal with properties of a single exciton. The chain emission is extended over 20 µm (for Fig. 1 b) which is likely the chain length. Its spectral profile does not vary along the fluorescence spatial extension (Fig. 1 c) proving that the potential confining the red exciton is highly regular, even on a macroscopic scale. Moreover, the excitation position neither affects the spectral intensity profile, nor the spatial one. Indeed, when changing the latter up to 10 µm along the chain, the fluorescence image remains identical (Fig. 1.d). These experimental facts show that the emission spatial profile is characteristic of the chain and is established very rapidly, on a timescale short compared to the 90 ps exciton effective lifetime [8].

To account for that, one can invoke conventional transport mechanisms for the exciton, like diffusive or ballistic transport. A diffusive transport of the exciton would require a diffusion coefficient $D = L^2/\tau \approx 10^4$ cm$^2$ s$^{-1}$, *L* being the fluorescence spatial extension (≈ 20 µm), and $\tau$ the effective radiative lifetime (≈ 100 ps at 10 K). This diffusion coefficient deduced to match the experimental data, is several orders of magnitude larger than values measured in organic or inorganic semiconductors [15]. A ballistic transport would require an exciton kinetic energy $E = \frac{1}{2} m_X^* \left(\frac{L}{\tau}\right)^2 \approx 10$ meV ($m_X^* \approx 0.1 m_0$ [8], $m_0$ being the free electron mass). However, at 10 K, the exciton average kinetic energy is of the order of 1 meV, being in thermodynamic equilibrium with the lattice [7]. Therefore neither of these two conventional transport mechanisms can explain the ultrafast macroscopic spatial extension of the fluorescence.

The delocalization of the red exciton center-of-mass over the entire chain is actually quite an attractive picture. In such a case, the exciton would be a quantum state, with a wave function having a macroscopic coherence length. In order to investigate this spatial coherence, an interference experiment was set up (see Fig. 1a and inset). The slits are used for selecting two spatially well separated emitting regions of the chain fluorescence and observe the interferences between them. This is the main difference with the classical Young's interference experiment where a single light source is split into two. High contrast interference fringes are observed, demonstrating extended spatial coherence of the radiating states over a length comparable to the total chain length. For a single 10 µm long chain, the resulting interference pattern is presented in Figure 2a-b. The fringes period scales like $\lambda f / a$ where $\lambda$ is the fluorescence wavelength, $f$ the focal length of the spectrometer entrance lens and $a$ the distance between the slits (see Fig. 2c).

The formation of a macroscopically coherent state needs further understanding. A probable microscopic scheme that we may invoke is the following. First of all, let us stress that in this system a strong exciton-photon coupling regime can be reached, i.e. the eigenstate is a polariton [16, 17, 18]. According to energy and momentum conservation requirements, only exciton states with wave vector near $k = 0$ are coupled to light. Taking into account that incident light is focused within an angle $\theta$ of 36°, the incident photons have a wave vector dispersion around $k_v = 0$ equal to $k_v \sin\theta / 2 \approx 5\times10^{-4}$ Å$^{-1}$ ($k_v = nE_v/\eta c$ is the photon wave vector in the material which refractive index $n = 1.5$ and $E_v = 2.28$ eV is the exciton transition energy). Thus, the excitons coupled to light have a wave vector along the chain axis such that $k \leq k_v \sin\theta/2$. If we compare to the edge $\pi/d$ of the first Brillouin zone, where $d$ is the monomer unit cell ($\approx 10$ Å) the wave vector of the photocreated exciton is very close to $k = 0$. Only excitons within an energy range $\Delta_{rad} = \eta^2 k_v^2 / 2m_X \approx 10$ μeV can then recombine radiatively assuming that all the lowest lying states in the band are thermally populated[19]. For a given chain length $L$ (10 μm in the case of the chain studied in Fig. 2) the accessible momentum states are quantized and separated by $\Delta k = \pi / L$ so that about 20 states are photocreated. Considering the initial excitation (at $x = 0$, $t = 0$), a coherent superposition of those different $k$-states, being in phase at $x = 0$, is created, leading to the formation of a state with a pronounced maximum at $x = 0$ and extending over the area of the exciting incident light ($\approx 1$ μm). This picture is similar to the formation of a quantum "wave packet". Then in a picosecond time range [7], the interaction with acoustical phonons dephase the different $k$-modes. Thus, the final state spreads over the total length of the chain and has an homogeneous broadening $\Gamma_h$ inversally proportional to this interaction time. This microscopic picture of the exciton state formation corresponds to a transient regime that cannot be observed in the cw experiments presented above. Therefore, what we do observe when averaging in time, is a stationary macroscopic quantum state of given energy, with a wave function of given phase and coherence length limited by the chain length. When probing the spatial coherence between two well-separated positions on the chain the dephasing is stationary in time and depends only upon the difference in the optical path, leading to an interference pattern like shown in Fig. 2a.

The fringes contrast can be very high, up to 75 % (Fig. 2b). However variations of the fringes constrast have been observed depending on the chosen positions on the chain. A possible explanation for contrast variations could lie in the non-uniform spatial emission profile that is usually observed (Fig. 1d). Indeed if the two emitting regions intensities going through the slits are different, then the interference contrast will be reduced. Actually, the spatial profile of the fluorescence can be influenced by local disorder present along the chain (mostly associated to elastic stress in the crystal). This would induce small amplitude potential fluctuations creating a non-uniform emission profile along the chain without destroying the spatial coherence of the state.

Acknowledgements

This work has been supported by the Region Ile de France (SESAME N°E. 1751).

Competing interests statement

The authors declare that they have no competing financial interests.

Correspondence and requests for materials should be addressed to V.V. (Valia.Voliotis@insp.jussieu.fr)

Figure legends

Figure 1: Microscopic imaging spectroscopy of a single chain.

**a.** Sketch of the micro-photoluminescence setup. A highly diluted PDA sample, with a concentration of red chains less than $10^{-8}$ in weight, is mounted on the cold finger of a Helium cryostat and cooled down to 10 K. Using a *f* =7.8 mm microscope objective with a 0.6 numerical aperture mounted on 3D piezo-actuators, a single isolated red chain is excited along 1 µm. The overall spectral and spatial resolutions are 50 µeV and 1 µm respectively. The chain excitonic fluorescence is collected by the same objective and either directly imaged on the nitrogen cooled CCD camera of an imaging spectrometer, or focused with a 800 mm lens on the slits for the interference experiment. The magnification is ≈ 100 while the slits have a 20 µm width and their separation can be varied, from 200 to 900 µm. Therefore two 1 µm regions of the chain, length limited by the spatial resolution, are selected and their separation can be tuned from 2 to 9 µm. The inset in the figure shows schematically how the chain fluorescence is imaged on the slits. $a$ is the variable distance between the two slits.

**b.** Spectrally resolved fluorescence image for a single isolated red chain. The vertical axis is the position along the chain in µm. The horizontal axis is the emission energy in eV. The excitation laser spot ($x$ = 0 in the text) is centered at the position 50 µm.

**c.** Normalized fluorescence spectral profile taken at 60 and 55 µm (grey circles and squares respectively). The full line corresponds to a lorentzian fit with a full width at half maximum of 510 µeV.

**d.** Normalized fluorescence spatial profile when moving the laser excitation spot along the chain. The resulting profiles are obtained for an excitation centered at 50, 55 and 60 µm for dots, triangles and squares respectively.

Figure 2 : Interference pattern of a single chain emission.

**a.** Interference pattern obtained from two 1µm wide emitting regions of a 10 µm long chain. This interference pattern is observed in the Fourier plane formed in the focal plane of the spectrometer entrance lens. The pattern length is limited by the collecting optics. The separation between the two sources taken out of the chain is about 2 µm. The corresponding distance in the plane of the slits is ≈ 220 µm leading to the fringes period of ≈ 225 µm.

**b.** Cross-section of the previous interference pattern at 2.2875 eV (see arrows in Fig 2.a) showing the intensity profile which gives a value of the contrast $C = (I_{max} - I_{min})/(I_{max} + I_{min})$ of 75%.

**c.** Plot of the fringes period as a function of the distance between the two interfering regions. The plot shows the results obtained for three experiments each performed on an individual chain (squares, circles and crosses). The fringe period scales like $\lambda f/a$ as shown by the straight line. The scaling factor $\lambda f$ only depends on the wavelength $\lambda$ of the emission and on the focal length $f$ of the spectrometer entrance lens.

**d.** Fourier transform of the cross section presented in b). A single spatial frequency appears corresponding to the fringes period.

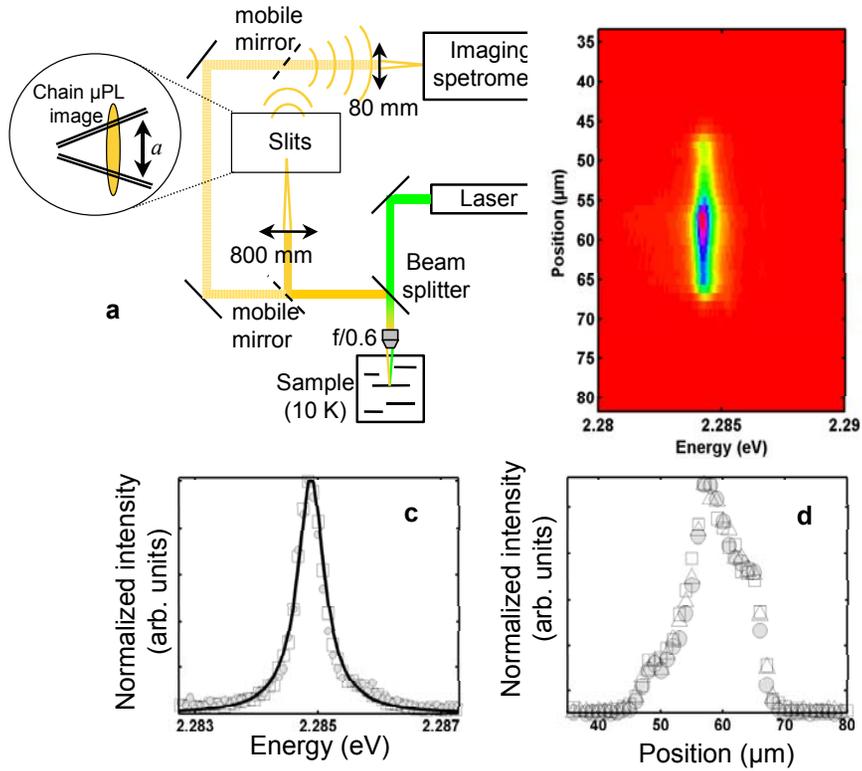

Figure 1: Microscopic imaging spectroscopy of a single chain.

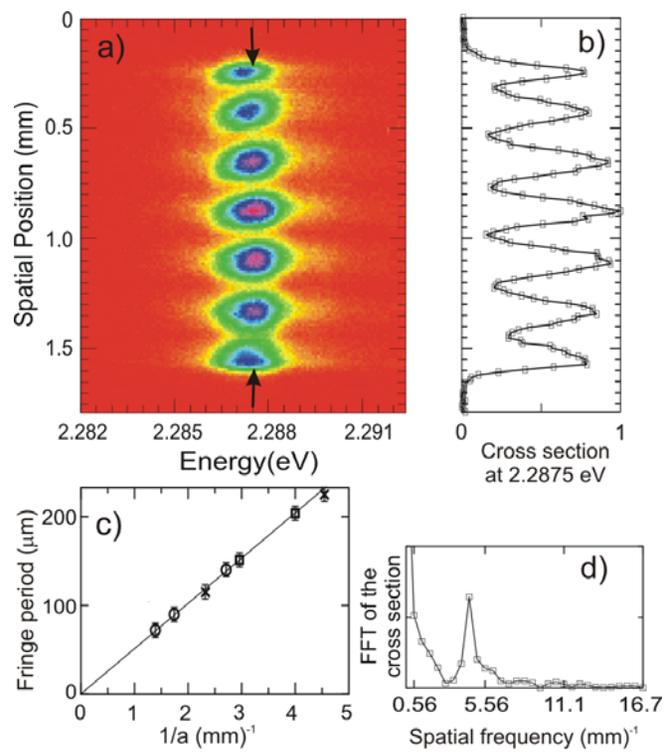

Figure 2 : Interference pattern of a single chain emission.